\documentclass[preprint]{aastex}

\def\s{{\rm s}}

\def\cm{{\rm cm}}
\def\km{{\rm\,km}}
\def\mm{{\rm\,mm}}
\def\gm{{\rm\,g}}

\def\K{{\rm\,K}}
\def\yr{{\rm\,yr}}

\def\dyne{{\rm\,dyne}}

\begin{document}
\shorttitle{Chondrules}
\shortauthors{Chiang}

\title{Chondrules and Nebular Shocks}

\author{E.~I.~Chiang}
\affil{Center for Integrative Planetary Sciences\\
Astronomy Department\\
University of California at Berkeley\\
Berkeley, CA~94720, USA}

\email{echiang@astron.berkeley.edu}

\begin{abstract}
Beneath the fusion-encrusted surfaces of the most primitive stony meteorites
lies not homogeneous rock, but a profusion of millimeter-sized
igneous spheres. These chondrules, and their
centimeter-sized counterparts, the calcium-aluminum-rich
inclusions, comprise more than half of the volume fraction of chondritic
meteorites. They are the oldest creations of the solar system.
Their chemical composition matches that of the solar photosphere
in all but the most volatile of elements, reflecting their
condensation from the same pristine gas that formed the sun.
In this invited editorial, we review the nebular shock wave model
of Desch \& Connolly (Meteoritics and Planetary Science 2002, 37, 183)
that seeks to explain their origin. While the model succeeds
in reproducing the unique petrological signatures of chondrules,
the origin of the required shock waves in protoplanetary disks
remain a mystery. Outstanding questions are summarized, with
attention paid briefly to competing models.
\end{abstract}


Beneath the fusion-encrusted surfaces of the most primitive stony meteorites
lies not homogeneous rock, but a profusion of millimeter-sized
igneous spheres [see Hewins (1996) and other articles in the
excellent compendium edited by Hewins, Jones, \& Scott].
These {\it chondrules}, and their
centimeter-sized counterparts, the {\it CAIs} (calcium-aluminum-rich
inclusions), comprise more than half of the volume fraction of chondritic
meteorites. They are the oldest creations of the solar system;
the oft-quoted age of the solar system of $4.566\pm 0.002$ billion years
refers to the crystallization ages of CAIs as determined from
radioactive isotope dating. Their chemical composition
matches that of the solar photosphere in all but the most volatile
of elements, reflecting their condensation from the same pristine
gas that formed the sun. Their petrology is consistent with their being heated
to super-liquidus temperatures for a period of a few minutes;
their roundness suggests that the heating occurred
while chondrule precursors were suspended in space,
so that surface tension pulled their shapes into spheres.
In the two hundred years since the discovery of chondrules,
this heating event has shrouded itself in secrecy. Identifying the
mechanism would be prize enough in itself; but the stakes are
potentially even greater for those who suspect that the primitive
character of chondrules and their substantial volume-filling fraction
implicate them in the equally mysterious process of planet
formation; that micron-sized dust grains could agglomerate to kilometer-sized
planetesimals only by first taking the form of millimeter-sized,
molten marbles; that the chondrule heating mechanism and the
means of planetesimal assembly are part and parcel of the same
physical process.

Numerous theories have been proposed for the formation of chondrules.
Many are staged within the primordial
solar nebula, the circumsolar disk of gas and dust from which the
planets congealed. None of these proposals has gained general acceptance.
The proposals (see the summary by Boss 1996)
range from nebular lightning (but how can we hope
to understand electrical discharges in the solar nebula when we
fail to understand them on Earth?), to collisions between molten
planetesimals, to irradiation by particle flares in the
vicinity of a magnetically active early sun (Shu et al. 2000).
Rarely are the predictions
of any given theory so detailed as to warrant comparison with the
wealth of experimental data available on chondrites. The article
by Desch and Connolly constitutes one of these welcome exceptions.
The potent combination of theoretical astrophysicist and
experimental petrologist consider the hypothesis
that chondrules were heated by shock waves propagating through the nebula.
Without devoting much attention to the question of the origin of these shocks,
they ask whether a nebular shock wave could {\it in principle}
generate thermal
histories for chondrules that are consistent with the mineralogical
and textural evidence (see also Hood \& Horanyi 1993; Connolly \& Love 1998).
The answer is enthusiastically affirmative.
Solving the equations of conservation of mass, momentum, and energy,
they compute the detailed temperature and density profile of a
one-dimensional, plane-parallel, steady shock. They conclude that
shock waves propagating at velocities of $v_s \sim 7 \km/\s$ through
gas having initially undisturbed temperatures of $T_1 \approx 300\K$,
densities $\rho_1 \approx 10^{-9}\gm/\cm^{3}$, and
chondrule concentrations of $10^{-8}$--$10^{-6}$ precursor particles/cm$^3$
(where a chondrule precursor is a ferromagnesian sphere of radius
0.3 mm and density $3\gm/\cm^3$) can reproduce empirically
determined thermal histories for chondrules.
These initial environmental parameters are chosen to resemble
those of standard models of protoplanetary disks at a heliocentric
distance of 2.5 AU [see, e.g., the minimum-mass solar nebula, obtained by
augmenting the masses of the planets to solar composition
and spreading that material in radius (Weidenschilling 1977).]
On its approach to the shock front, the precursor is heated
to temperatures of $\sim$1500 K (just below the liquidus)
for $\sim$1 hour by absorbing
radiation emitted by yet hotter chondrules and nebular dust
that are further downstream past the front. Immediately after
crossing the front, the precursor encounters a supersonic
headwind of gas and is frictionally heated to temperatures of $\sim$1800 K
for a few minutes. Subsequent cooling is slowed by the fact that
chondrules remain in thermal contact with hot shocked gas. The time
to cool through solidus is given by the time the chondrule takes
to travel several optical depths away from the intensely luminous
shock front; this cooling timescale is several hours for the parameters above,
in accord with experiment. Increasing the precursor concentration
enhances the degree of pre-shock radiative heating
and thereby increases the peak temperatures that chondrules attain.
Since greater chondrule concentrations imply greater rates of collisions
among them, and since higher peak temperatures give rise to radial
or barred textures as opposed to
porphyritic textures in chondrules synthesized in the lab,
the model predicts the incidence of
compound chondrules (two or more chondrules bound together
by a collision in which one or more of the
bodies was still plastic)
to be higher among radial/barred chondrules than among porphyritic ones.
This correlation is, in fact, observed in nature.

Can shocks explain naturally the fact that
chondrules have nearly uniform sizes of $\sim$1 millimeter? Desch and Connolly
defer this question to future work but we may guess that the
answer is positive here as well. The maximum size of a chondrule
would be set by the same physics that sets the size of
a raindrop (P. Goldreich 1997). By balancing
the cohesive force of surface tension against the destructive
force of turbulent gas drag, we estimate a maximum chondrule radius
of $\sim$4$\gamma/\rho_2 v_s^2 \sim 4\mm$, where
$\rho_2 \approx 10^{-8} \gm/\cm^{3}$
is the density just past the shock front and $\gamma \approx 500 \dyne/\cm$
is the surface tension of molten rock. Molten droplets greater than this size
would bifurcate as they plow supersonically
through shocked gas. The minimum size would be set by evaporation
in the post-shock flow;
particles of radius $\lesssim 0.1\mm$ sublimate away
if kept at temperatures of $\sim$1700 K for a few hours.

Nebular shocks may even provide a means of agglomerating chondrules
after heating them. At post-shock distances greater than those considered
by Desch and Connolly, we expect chondrules and gas to cool
back down to their initial temperatures of $\sim$300 K. The jump
conditions for a one-dimensional, plane-parallel isothermal shock
yield a final post-shock gas density $\rho_3$ that is greater than
$\rho_1$ by a factor of order $mv_s^2/kT_1 \sim 10^2$, where
$m$ is the mass of a hydrogen molecule and $k$ is Boltzmann's constant.
For standard parameters, $\rho_3 \sim 10^{-7}\gm/\cm^3$---great
enough in the context of standard nebular models at heliocentric
distances of 2.5 AU that matter may clump together under its
own self-gravity.

There is, however, an entire other dimension to the experimental
data that the mechanism of nebular shocks does not address: the
presence of once-active, short-lived radionuclides such as
$^{26}$Al in CAIs (Lee et al.~1976). The long-standing hypothesis that these
radionuclides were produced in supernovae that externally seeded
the solar nebula has been called into question by the discovery
of $^{10}$ Be in CAIs (McKeegan et al.~2000), since $^{10}$Be is
not a stellar nucleosynthetic product. The recent competing theory
of Shu et al.~(2000) and Gounelle et al.~(2001) that chondrules
were irradiated by particle flares in the vicinity of an active early Sun
offers a framework for understanding the origin of radioactive nuclides
in CAIs, including $^{10}$Be. It is not clear, however, that
the latter theory can account for the extensive petrographic data on
which Desch and Connolly focus.

Finally, what is the origin of nebular shocks? Desch and Connolly
espouse self-gravitating clumps of matter (the potential
building blocks of proto-Jupiter) that orbit at $\sim$5 AU
and that gravitationally drive density structures within
the chondrule forming region at $\sim$2.5 AU.
This and other proposals are not sufficiently developed
to predict the number of times a given chondrule passes
through shock fronts. Empirically, the number of times
a chondrule is heated must
be between 1 and $\sim$3, based on observations of chondrule rims
(Rubin \& Krot 1996; Hewins 1996).
The proposal for the origin of shock waves by Desch and Connolly may
run aground on this point---if the massive bodies at $\sim$5 AU
are present for the lifetime of the
nebula, $T \sim 10^6\yr$, then shock fronts will have processed
material along the entire circumference of the asteroid belt
$T/T_{con} \sim 3 \times 10^4$ times, where
$T_{con} \sim 30\yr$ is the time between
conjunctions of a body at 5 AU and a body at 2.5 AU.
While Desch and Connolly provide useful, state-of-the-art
computations of the thermal histories of particles
traversing shock fronts, until a convincing source of shock waves
is identified, the problem of chondrule formation will
remain unsolved at the zeroth order level.

\end{document}